\documentclass[a4paper, 11pt]{article}
\usepackage[paperwidth=8.5in,paperheight=11in,top=1in, bottom=1in, left=1in, right=1in]{geometry}

\usepackage{amsmath,amssymb,amsthm,amsfonts,cases}
\usepackage{mathrsfs}
\usepackage{url}
\usepackage[dvipsnames]{xcolor}
\usepackage{graphicx}
\usepackage[authoryear]{natbib}
\usepackage{setspace}

\usepackage{authblk}

\allowdisplaybreaks

\usepackage{hyperref} 
\hypersetup{
	colorlinks   = true,
	citecolor    = blue,
	linkcolor    = OrangeRed,
	urlcolor     = magenta
}

\theoremstyle{plain}
\newtheorem{theorem}{Theorem}[section]
\newtheorem{lemma}[theorem]{Lemma}

\newtheorem{corollary}[theorem]{Corollary}

\newtheorem{problem}{Problem}

\newtheorem{example}[theorem]{Example}
\newtheorem{remark}[theorem]{Remark}

\DeclareMathOperator*{\argmin}{arg\,min}

\makeatletter\@addtoreset{equation}{section} \makeatother

\newcommand{\Eb}{\mathbb{E}}
\newcommand{\Fb}{\mathbb{F}}
\newcommand{\Pb}{\mathbb{P}}
\newcommand{\Qb}{\mathbb{Q}}
\newcommand{\Rb}{\mathbb{R}}
\newcommand{\Vb}{\mathbb{V}}

\newcommand{\Ac}{\mathcal{A}}
\newcommand{\Bc}{\mathcal{B}}

\newcommand{\Fc}{\mathcal{F}}

\newcommand{\Lc}{\mathcal{L}}

\newcommand{\dd}{\mathrm{d}}
\newcommand{\ee}{\mathrm{e}}

\newcommand{\wt}{\widetilde}

\newcommand{\Proj}{\mathrm{Proj}}

\usepackage{bbm}

\newcommand{\Pis}{\Pi_\sigma}
\newcommand{\xic}{\xi_{\mathrm{c}}}
\newcommand{\VQ}{V_{\mathrm{QLM}}}

\begin{document}

\title{\vspace{-2cm} Cone-constrained Monotone Mean-Variance Portfolio Selection\\Under Diffusion Models}

\author[a]{Yang Shen}
\author[b]{Bin Zou}

\affil[a]{\footnotesize School of Risk and Actuarial Studies, University of New South Wales, Sydney, NSW 2052, Australia. Email: y.shen@unsw.edu.au.}
\affil[b]{\footnotesize Corresponding author. 341 Mansfield Road U1009, Department of Mathematics, University of Connecticut, Storrs 06269-1009, USA. Email: bin.zou@uconn.edu.}

\date{\small First version: March 29, 2022; this version: \today\\
Forthcoming in \emph{SIAM Journal on Financial Mathematics}}
\maketitle

\begin{abstract}
\noindent
We consider monotone mean-variance (MMV) portfolio selection problems with a conic convex constraint under diffusion models, and their counterpart problems under mean-variance (MV) preferences. 
We obtain the precommitted optimal strategies to both problems in closed form and find that they coincide, without and with the presence of the conic constraint. 
This result generalizes the equivalence between MMV and MV preferences from non-constrained cases to a specific constrained 
case. A comparison analysis reveals that the orthogonality property under the conic convex set is a key to ensuring the
equivalence result. 
\end{abstract}





\section{Introduction}
\label{sec:intro}

Mean-variance (MV) preferences are popular optimization criteria in portfolio selection problems, dating back to the seminal work of \cite{markowitz1952portfolio}, and can be represented in the following form (see, e.g., \cite{li2000optimal} and \cite{zhou2000continuous}):
\begin{align}
	\label{eq:MV}
	\wt J_\theta(X) = \Eb^\Pb[X] - \frac{\theta}{2} \Vb^\Pb[X],
\end{align}
where $X \in \Lc^2(\Pb)$ denotes a risky position (e.g., the terminal value of a portfolio) that is square-integrable under some physical probability measure $\Pb$, $\Eb^\Pb$ and $\Vb^\Pb$ stand for the expectation and variance operators under $\Pb$, and $\theta > 0$ is a risk aversion parameter, characterizing the trade-off between mean (performance) and variance (risk). 
A major drawback of MV preferences is the lack of monotonicity, i.e., given $X \ge X'$ $\Pb$-a.s.,  MV preferences \eqref{eq:MV} may yield $\wt J_\theta(X) < \wt J_\theta(X')$ for some $\theta >0$ (see \cite{maccheroni2009portfolio}, p.488). In fact, \cite{cui2012better} show that it is even possible to improve the performance of a precommitted optimal MV portfolio by extracting a free cash flow stream (FCFS) in a discrete-time model.\footnote{Throughout this paper, we are only concerned with the so-called precommitted optimal strategies under MV and MMV preferences, which are time-inconsistent. For precommitted MV problems, please see  \cite{li2000optimal} for discrete-time models and \cite{zhou2000continuous} for continuous-time models. For time-consistent MV problems, we refer to \cite{basak2010dynamic} and \cite{bjork2010general} for earlier contributions. \cite{shen2021mean} provide a detailed comparison of these two type of optimal/equilibrium strategies.}

To address the non-monotonicity drawback of MV preferences, \cite{maccheroni2009portfolio} propose the so-called monotone mean-variance (MMV) preferences, defined by\footnote{See \cite{vcerny2020semimartingale} for an alternative characterization.} 
\begin{align}
	\label{eq:MMV}
	J_\theta (X) = \inf_{\Qb \in \Delta^2(\Pb)} \, \Eb^\Qb \left[X + \frac{1}{2\theta} \left(\frac{\dd \Qb}{\dd \Pb} - 1\right)\right],
\end{align}
where 
$\Delta^2(\Pb)$ denotes the set of all probability measures that have a square-integrable density with respect to $\Pb$
and $\theta > 0$ is a risk aversion parameter.
Although \eqref{eq:MMV} does not immediately reveal any direct link between MMV and MV preferences, \cite{maccheroni2009portfolio} show that (1) $J_\theta(X) = \wt J_\theta (X)$ if $X \in \mathcal{G}_\theta := \{X \in \Lc^2(\Pb): X - \Eb^\Pb[X] \le 1/\theta \}$, i.e., the monotone region\footnote{By Lemma 2.1 in \cite{maccheroni2009portfolio},  the region where the Gateaux differential of $J_\theta(X)$ is positive is given by $\mathcal{G}_\theta$; see Appendixes A and B therein for proofs.}
and (2) $J_\theta$ is the smallest monotone functional that dominates $\wt J_\theta$ outside of $\mathcal{G}_\theta$.

Upon verifying the desirable features of MMV preferences, \cite{maccheroni2009portfolio} solve MMV portfolio selection problems (MMV problems hereafter) in a single-period model for the first time in literature. 
An important result is that the optimal MMV strategy is \emph{different} from the optimal strategy to MV portfolio selection problems (MV problems hereafter). 
\cite{vcerny2012computation} establish an interesting relation between MMV preferences and truncated quadratic utility, and obtain an equivalence result between their associated portfolio selection problems.
\cite{vcerny2020semimartingale} show that an FCFS does not exist if MV problems and MMV problems have the same optimal solution. 
As such,  the result of \cite{maccheroni2009portfolio} is immediately consistent with the existence of an FCFS in \cite{cui2012better}, where the authors also note that FCFS extraction is \emph{not} possible in a continuous-time complete diffusion market model. 
 \cite{bauerle2015complete} further show that such a result holds true for any \emph{complete} markets. 
 \cite{trybula2019continuous} consider MMV problems under a stochastic factor model, which is an incomplete market model and driven purely by Brownian motions, and by applying the Hamilton–Jacobi–Bellman–Isaacs (HJBI) equations approach, they obtain the optimal MMV strategy and show that it is also optimal to MV problems. In a subsequent work, \cite{strub2020note} prove that the optimal strategies to MV problems and MMV problems always coincide for any market model with \emph{continuous} price processes, rendering the result of \cite{trybula2019continuous} as a special case. 
The intrinsic relation between MMV and MV preferences partially explains the scarcity of investigations on MMV problems; the work of \cite{trybula2019continuous} is the only one, to our knowledge, that directly tackles MMV problems in a continuous-time model and obtains the optimal MMV strategy in an explicit form. 
However, all the existing works reviewed above study the connection between the two preferences when investment strategies are \emph{unconstrained}.\footnote{\cite{cui2017mean} consider MV problems for convex cone-constrained discrete-time markets and study the relation between time consistency in efficiency (TCIE), proposed in \cite{cui2012better}, and the minimum-variance signed supermartingale measure (VSSM). In particular, \cite{cui2017mean} establish sufficient and necessary conditions of TCIE via the VSSM.}
It is unclear whether MMV problems will still yield the same optimal strategy as MV problems when portfolio constraints are present in continuous-time models.

This paper aims to take a first step towards the understanding of the relation between MMV problems and MV problems under portfolio constraints. 
To achieve this goal, we follow the same direction as \cite{trybula2019continuous}, i.e., we solve constrained MMV problems and constrained MV problems explicitly in a specific model and compare their optimal solutions to see whether they coincide. This approach differs significantly from \cite{vcerny2012computation}, \cite{strub2020note}, and \cite{vcerny2020semimartingale}, which analyze similar (or related) problems in a general setting but do not attempt to find optimal strategies (at least not in an easily implementable form). 
A strong motivation of taking this direction is the intrinsic difference between these two problems  (despite having the same solution in many cases) from the stochastic control viewpoint.
MV problems are a special case of stochastic linear-quadratic control problems, while MMV problems can be seen as a type of stochastic differential games (see \cite{mataramvura2008risk}) or robust optimization problems (see \cite{hansen2001robust}). 
Therefore, solving MMV problems is of independent interest in control theory, and indeed requires advanced mathematical tools to find the optimal strategy in a complex model (see \cite{fan2022monotone} in a non-Markovian regime-switching model).

With the above discussion in mind, we consider a simple pure diffusion model with constant parameters,\footnote{It is certainly possible to consider problems in a more general setting, but explicit solutions are likely unavailable or a much more involved analysis is required. For instance, \cite{li2002dynamic} consider MV problems under the no-shorting constraint in a similar model but with parameters given by deterministic functions, and need to resort to viscosity solutions to identify the optimal MV strategy.}  
consisting of one risk-free asset and $n$ risky assets, whose price processes are driven by $d$ Brownian motions, where $d \ge n$ (thus the market is possibly incomplete). 
By formulating constrained MMV problems as a stochastic differential game, we apply the HJBI equation approach to obtain the optimal MMV strategy in closed form (see Theorem \ref{thm:MMV}). 
We also solve the counterpart constrained MV problems in the same market (see Theorem \ref{thm:MV}), by utilizing the so-called embedding technique (see  \cite{zhou2000continuous}).
Through a comparison study, we find that the two optimal strategies coincide, either without or with the constraint. 
Therefore, on the one hand, our result verifies those in the literature (see, e.g., \cite{vcerny2012computation} and 
\cite{strub2020note}) in a specific model,  when constraints are not present; on the other hand, more importantly, ours is the first to reveal that the same 
equivalence between MMV and MV preferences holds in the presence of conic convex constraints, 
at least under diffusion market models. We also discover that a key to ensuring the equivalence result is the orthogonal 
property under the conic convex set (see Lemma \ref{lem:xi}), which is not necessarily true for a general convex set (see 
Remark \ref{rmk:general}). This 
observation leaves an 
interesting open question, i.e., whether the same equivalence result between MMV and MV preferences remains valid for 
general convex constraints.

We end this section with a detailed comparison with our working paper \cite{fan2022monotone}, which also studies constrained MMV problems, but under non-Markovian models and general convex constraints that are not necessarily conic. A strong contribution of \cite{fan2022monotone} is on the technical side; it employs complex tools, from stochastic Hamilton-Jacobi-Bellman equations and backward stochastic differential equations theory, to solve a robust portfolio optimization problem with constraints under non-Markovian (path-dependent) models. 
Here, since all model parameters are constants in the current paper, we are able to apply the standard HJBI method to obtain the optimal strategy; as such, the two papers differ greatly in terms of methodologies.	Preliminary results in \cite{fan2022monotone} show that, under non-conic constraints, the optimal strategy under MMV preferences has a \emph{different} expression from the one under MV preferences. 
However, due to the complexity of the model and problem in \cite{fan2022monotone}, we are unable to simplify the two optimal strategies and provide  a counterexample to show that they are different at the moment. For this very reason, we decide to consider a tractable diffusion model to investigate the impact of constraints on the relation between MMV preferences and MV preferences. At first sight, the optimal MMV strategy $\pi^*(\cdot)$ in \eqref{eq:can} and the optimal MV strategy $\wt \pi^*(\cdot)$ in \eqref{eq:pi_MV} look very different in this current paper; but upon recognizing the ``hidden'' conditions \eqref{eq:relation} and \eqref{eq:rela_xi}, we show that the two actually coincide, with and without constraints (see Corollaries \ref{cor:un} and \ref{cor:con}). Furthermore, the comparison analysis in Section \ref{sub:com} shows that \emph{conic} convex constraints only contribute to minor changes in the values of the optimal strategies, but do not alter the main structure (please compare \eqref{eq:pi_un} and \eqref{eq:form}), which in turn preserves the equivalence under the constrained case.

The rest of the paper is organized as follows. In Section \ref{sec:model}, we formulate constrained MMV problems (Problem 
\ref{prob:MMV}) and constrained MV problems (Problem \ref{prob:MV}). 
Section \ref{sec:main} contains all the main results, including explicit solutions to both Problems \ref{prob:MMV} and 
\ref{prob:MV} and a detailed comparison study that validates the two optimal strategies coincide when conic convex 
constraints are present. Finally, Section \ref{sec:con} concludes the paper.

\section{Model}
\label{sec:model}

Let us fix a probability space $(\Omega, \Fc, \Fb := (\Fc_t)_{t \in [0,T]}, \Pb)$ over a finite horizon $[0,T]$, where 
the filtration $ \Fb$ is generated by a $d$-dimensional standard Brownian motion $W(\cdot) := (W_1(\cdot), W_2(\cdot), 
\ldots, W_d(\cdot))^\top$, with $\top$ denoting the usual transpose operation.
We consider a continuous-time frictionless financial market with one risk-free asset, which earns interest at a constant rate $r \in \Rb$, and $n$ risky assets, whose price processes $S_i(\cdot)$ are given by the following geometric Brownian motion (GBM) model:
\begin{align*}
	\dd S_i(t) = S_i(t) \Big( \mu_i \, \dd t + \sum_{j=1}^d \sigma_{ij} \, \dd W_j(t) \Big), \qquad S_i(0)>0, \, i=1,2, \ldots, n,
\end{align*}
where $\mu_i$ and  $\sigma_{ij}$ are positive constants. 
Let the $n$-dimensional vector $\mu:=(\mu_1, \mu_2, \ldots, \mu_n)^\top$ denote the expected returns, $B:=\mu - r \mathbf{1}$ the excess return vector, and $\sigma := (\sigma_{ij})$ the $n \times d$ volatility matrix. 
We assume that $d \ge n$ (the financial market is possibly incomplete) and $\Sigma := \sigma \sigma^\top$ is positive definite (implying its inverse $\Sigma^{-1}$ exists and the underlying market is arbitrage-free). 

We consider portfolio selection problems in the above-described financial market with self-financing strategies $\pi(\cdot) := (\pi_1(\cdot), \pi_2(\cdot), \ldots, \pi_n(\cdot))^\top$, where $\pi_i(\cdot) := \{\pi_i(t)\}_{t \in [0,T]}$ denotes the investment \emph{amount} in the $i$-th risky asset over $[0,T]$. 
Given a strategy $\pi(\cdot)$, the associated wealth process $X(\cdot) := \{X(t)\}_{t \in [0,T]}$ is governed by the following stochastic differential equation (SDE):
\begin{align}
	\label{eq:dX}
	\dd X(t) = \big(r X(t) + \pi(t)^\top B \big) \dd t + \pi(t)^\top \sigma \, \dd W(t), \quad X(0) = x_0 \in \Rb.
\end{align}
We model portfolio constraints by a conic convex  set $\Pi \subset \Rb^n$.
Such a formulation covers several interesting cases: (1) if we take $\Pi = \Rb_+^n$,  then short selling any risky assets is not allowed; (2) if we take $\Pi = \{0\} \times \Rb^{n-1}$, then the first risky asset is non-tradable. 
Letting $\Pis := \{\sigma^\top \pi: \pi \in \Pi\}$, we easily see that 
$\pi(t) \in \Pi \Leftrightarrow \sigma^\top \pi(t) \in \Pis$, where $t \in [0,T]$. Thus, $\Pis$ is a conic convex set in $\Rb^d$.
Define the following operator $\Proj: \Rb^d \mapsto \Pis$:
\begin{align}
	\label{eq:Proj}
	\Proj_{\Pis} [v] := \argmin_{v' \in \Pis} \, \left\|v - v' \right\|^2, \quad v \in \Rb^d,
\end{align}
which is the \emph{orthogonal projection} of $v \in \Rb^d$ onto the set $\Pis \subset \Rb^d$. 
The conic assumption of $\Pi$ implies that $\Proj_{\Pis} [k v] = k \, \Proj_{\Pis} [v]$ for any $k \in \Rb$ and any $v \in \Rb^d$ (i.e., the operator $\Proj_{\Pis}$ is homogeneous with degree 1).

To formulate constrained portfolio selection problems, we first introduce the concept of admissible investment strategies.
An investment strategy $\pi(\cdot)$ is called admissible if (1) it is $\Fb$-progressively measurable and square-integrable under $\Pb$; (2) $\pi(t) \in \Pi$ for all $t \in [0, T]$; (3) the SDE \eqref{eq:dX} has a unique strong solution $X(\cdot)$ such that $\Eb^\Pb[\sup_{t \in [0,T]} |X(t)|^2 ] < +\infty$. 
Denote the set of all admissible strategies by $\Ac$.

To characterize $\Qb$ in the MMV preferences \eqref{eq:MMV}, we follow \cite{trybula2019continuous} and introduce a distortion process $\eta(\cdot)$ to define $\Qb$ as an equivalent measure to $\Pb$ via the change of measure technique.\footnote{This technique is also popular in the modeling of alternative probability measures in robust portfolio optimization (ambiguity) literature (see, e.g., the influential work of \cite{hansen2001robust}).} 
To be precise, given a suitable distortion process $\eta(\cdot)$, we define $\Qb = \Qb^\eta$ by
\begin{align}
	\label{eq:dQ}
	\frac{\dd \Qb^\eta}{\dd \Pb}\bigg|_{\Fc_t} :=\Lambda^\eta(t) = \exp\left(\int_0^t \, \eta(s)^\top \, \dd W(s) - \frac{1}{2} \int_0^t \left\| \eta(s) \right\|^2 \, \dd s\right), \quad t \in [0,T].
\end{align}
To ensure that $\Qb^\eta$ is well defined by \eqref{eq:dQ}, we assume that the process $\eta(\cdot)$ satisfies the following conditions: (1) it is $\Fb$-progressively measurable;
(2) 
\begin{align*}
	\Eb^\Pb \left[ \big(\Lambda^\eta(t)\big)^2 \right] < +\infty \quad \text{ and } \quad \Eb^\Pb[\Lambda^\eta(t)] = 1, \quad t \in [0,T].
\end{align*}
(Please see Section 2.1 of \cite{trybula2019continuous} for similar conditions.)
By condition (2), the process $\Lambda^\eta(\cdot)$ is a true martingale. 
Denote $\Bc$ the set of all processes that meet the above two conditions. 
Denoting $W^{\Qb^\eta}(\cdot):=\{W^{\Qb^\eta}(t)\}_{t \in [0,T]}$ {a} $d$-dimensional $\Qb^\eta$-Brownian motion, we have 
\begin{align*}
	 \dd W(t) = \dd W^{\Qb^\eta}(t) + \eta(t) \dd t, \quad t \in [0,T].
\end{align*}

We are now ready to formulate constrained MMV problems as follows:

\begin{problem}[Constrained MMV Problems]
	\label{prob:MMV}
	We seek an optimal pair of strategies $(\pi^*(\cdot), \eta^*(\cdot))  \in \Ac \times \Bc$ that optimizes the MMV preferences \eqref{eq:MMV} over all admissible strategies in $\Ac \times \Bc$.
	The value function of the dynamic version of the MMV problems, $V: [0,T] \times \Rb \times \Rb_+ \mapsto \Rb$,  is 
	defined by\footnote{In \eqref{eq:Vt}, both $\Ac$ and $\Bc$ are understood as the dynamic version defined over $[t, T]$.} 
	\begin{align}
		\label{eq:Vt}
		V(t,x,\lambda) := \sup_{\pi(\cdot) \in \Ac} \, \inf_{\eta(\cdot) \in \Bc} \, \Eb^{\Qb^\eta}_{t,x,\lambda} \left[X(T) + \frac{1}{2\theta} \left( \Lambda^\eta(T) - 1\right)\right], \quad \theta >0,
	\end{align}
	where $\Eb^{\Qb^\eta}_{t,x,\lambda}$ denotes taking conditional expectation given $X(t) = x \in \Rb$ and ${\Lambda^\eta}(t) = \lambda \in \Rb_+$. 
	We call $\pi^*(\cdot)$ the optimal investment strategy and $\eta^*(\cdot)$ the optimal distortion process.
\end{problem}

As discussed in Section \ref{sec:intro}, MMV problems often have an interesting relation with their MV counterpart problems, which are formulated in the following:

\begin{problem}[Constrained MV Problems]
	\label{prob:MV}
We seek an optimal strategy $\wt \pi^*(\cdot) \in \Ac$ that maximizes the MV preferences \eqref{eq:MV} over all admissible strategies in $\Ac$. Denoting $\wt X^*(\cdot)$ the wealth process \eqref{eq:dX} under $\wt \pi^*(\cdot)$, the  value function is given by 
\begin{align*}
	\wt V (x_0) := \wt J_\theta(\wt X^*(T)) = \sup_{\pi(\cdot) \in \Ac} \, \left\{ \Eb^\Pb[X(T)] - \frac{\theta}{2} \Vb^\Pb[X(T)] \right\}, \quad \theta>0.
\end{align*}
\end{problem}

\section{Main Results}
\label{sec:main}

We solve the constrained MMV problems in Section \ref{sub:MMV} and the constrained MV problems in Section \ref{sub:MV}, and conduct a detailed comparison study on their optimal strategies in Section \ref{sub:com}.

\subsection{Solution to Constrained MMV Problems}
\label{sub:MMV}

This section offers a complete solution to Problem \ref{prob:MMV} (constrained MMV problems), with detailed results summarized in Theorem \ref{thm:MMV}.
To achieve this objective, we consider the dynamic version of Problem \ref{prob:MMV}  as defined in \eqref{eq:Vt}. 
Under suitable regularity conditions,
the value function $V=V(t,x,\lambda)$ satisfies the following Hamilton-Jacobi-Bellman-Isaacs (HJBI) equation:
\begin{align}
	0 = \sup_{\pi \in \Pi} \inf_{\eta \in \Rb^d} \Big\{ \underbrace{V_t + V_x \big( rx + \pi^\top (B + \sigma \eta) \big) + V_\lambda \lambda \left\|\eta\right\|^2   + \frac{1}{2} V_{xx} \pi^\top \Sigma \pi + \frac{1}{2} V_{\lambda \lambda} \lambda^2  \left\|\eta\right\|^2 + V_{x \lambda} \lambda \pi^\top \sigma \eta}_{:=\Lc^{\pi, \eta}V(t,x,\lambda)} \Big\},  
\label{eq:HJBI}
\end{align}
along with the boundary condition
\begin{align}
	\label{eq:boundary}
	V(T, x, \lambda) = x + \frac{1}{2 \theta} (\lambda - 1), \qquad x \in \Rb, \lambda \in \Rb_+.
\end{align}
In \eqref{eq:HJBI}, $V_\cdot$ denotes the corresponding partial derivative (e.g., $V_t = \frac{\partial V}{\partial 
t}(t,x,\lambda)$) and $\Sigma = \sigma \sigma^\top$.

Lemma \ref{lem:veri} below provides a verification theorem for Problem \ref{prob:MMV}, similar to Theorem 1 in \cite{trybula2019continuous}, whose proof is omitted here (see Theorem 3.2 and its proof in \cite{mataramvura2008risk} for a standard reference).
As such, the key of solving Problem \ref{prob:MMV} is finding a solution to \eqref{eq:HJBI}-\eqref{eq:boundary}. 

\begin{lemma}
\label{lem:veri}
Suppose there exists a function $\phi \in \mathrm{C}^{1,2,2}((0,T) \times \Rb \times (0,\infty)) \cap \mathrm{C}([0,T] \times \Rb \times [0,\infty) )$ and an admissible Markov control $(\pi^*(\cdot), \eta^*(\cdot)) \in \Ac \times \Bc$ such that
(i) $\Lc^{\pi^*(t), \eta} \phi(t,x,\lambda) \ge 0 $;
(ii) $\Lc^{\pi, \eta^*(t)} \phi(t,x,\lambda) \le 0 $;
(iii) $\Lc^{\pi^*(t), \eta^*(t)} \phi(t,x,\lambda) = 0$;
(iv) $\phi(T,x,\lambda) = x + \frac{1}{2\theta} (\lambda - 1)$,
for all $\pi \in \Pi$, $\eta \in \Rb^d$, and $(t,x,\lambda) \in [0,T) \times \Rb \times \Rb_+$; 
and (v)
\begin{align*}
	\Eb_{t,x,\lambda}^{\Qb^\eta} \left[\sup_{s \in [t, T]}  \, \big| \phi(s, X(s), \Lambda^\eta(s)) \big| \right] < + \infty
\end{align*}
for all $\pi(\cdot) \in \Ac$, $\eta(\cdot) \in \Bc$, and $(t,x,\lambda) \in [0,T) \times \Rb \times \Rb_+$. 
Then $\phi(t,x,\lambda) = V(t,x,\lambda)$, where $V$ is defined in \eqref{eq:Vt}, and $(\pi^*(\cdot), \eta^*(\cdot))$ is an optimal pair of strategies to Problem \ref{prob:MMV}.
\end{lemma}

\begin{theorem}
	\label{thm:MMV}
 The value function $V(t, x, \lambda)$ defined in \eqref{eq:Vt} is given by 
\begin{align}
	\label{eq:V_ansatz}
	V(t,x,\lambda) = x h(t) + \frac{1}{2 \theta} \big(\lambda f(t) - 1 \big), \qquad t \in [0,T], x \in \Rb, \lambda \in \Rb_+,
\end{align}
where the deterministic functions $h$ and $f$ are defined respectively by 
\begin{align}
	\label{eq:h_f}
	h(t) &:= \ee^{r(T-t)}  \quad \text{ and } \quad f(t) := \ee^{\left( 2 \xi^\top \xic - \left\| \xic \right\|^2 \right) (T-t)}, \quad t \in [0,T],
	\\
	\label{eq:xi}
\text{with} \quad \xi &:= \sigma^\top \Sigma^{-1} B \quad \text{ and } \quad 	\xic := \Proj_{\Pis} \big[ \sigma^\top \Sigma^{-1} B \big] = \Proj_{\Pis} \big[\xi \big].
\end{align}
The optimal investment strategy	$\pi^*(\cdot):=\{\pi^*(t)\}_{t \in [0,T]}$ to  Problem \ref{prob:MMV} is given by
\begin{align}
	\pi^*(t) = - \left(X^*(t) - x_0 \ee^{rt} - \frac{1}{\theta} \, \ee^{-r(T-t)} \big(f(0) + 
	G(0,t) \big) \right)\, \Sigma^{-1} \sigma \, \xic, \quad t \in [0,T],
	\label{eq:pi_MMV}
\end{align}
where $X^*(\cdot)$ satisfies \eqref{eq:dX} under the above $\pi^*(\cdot)$ and $X^*(0) = x_0$, $f(0)$ and $\xic$ are given by \eqref{eq:h_f}-\eqref{eq:xi}, and $G(0, t)$ is defined in \eqref{eq:G} below.
The optimal distortion process $\eta^*(\cdot):=\{\eta^*(t)\}_{t \in [0,T]}$ is given by
\begin{align}
	\eta^*(t) = -\Proj_{\Pis} \big[ \sigma^\top \Sigma^{-1} B \big]  =- \xic.
	\label{eq:eta_MMV}
\end{align}

\end{theorem}

\begin{proof} Based on the boundary condition \eqref{eq:boundary}, we guess an {\it ansatz} of the value function in the form of 
\eqref{eq:V_ansatz}, where the two deterministic functions $h$ and $f$ are yet to be determined and satisfy $h(T) = f(T) = 
1$.
In addition, we assume that both $h$ and $f$ are differentiable and  $h(t) > 0$ and $f(t) > 0$ for all $t \in [0,T]$, which guarantee that the first-order condition is also sufficient for both optimization problems in the HJBI equation \eqref{eq:HJBI}. 
From the linear ansatz form, we easily see that $V_{xx} = V_{x\lambda} = V_{\lambda \lambda} =0$.

Given the above ansatz $V$ and any $\pi \in \Pi$, solving the infimum problem over $\eta \in \Rb$ in \eqref{eq:HJBI} yields 
\begin{align*}
	\eta^*(\pi) = - \frac{\theta \, h}{\lambda \, f} \, \sigma^\top \pi.
\end{align*}
By plugging the above $\eta^*(\pi)$ back into \eqref{eq:HJBI} and applying the first-order condition to the supremum problem over $\pi \in \Rb^n$ (unconstrained), we establish the following equation:
\begin{align*}
	\sigma^\top \pi = \frac{\lambda \, f}{\theta \, h} \, \sigma^\top \Sigma^{-1} B,
\end{align*}  
which implies that the solution to the constrained problem over $\pi \in \Pi$ is  given by
\begin{align}
	\label{eq:can}
	\pi^* = \frac{\lambda \, f}{\theta \, h} \, \Sigma^{-1} \sigma \, \xic, \quad \text{ and thus } \quad \eta^* = \eta^*(\pi^*) = - \xic,
\end{align}
where $\xic$ is defined in \eqref{eq:xi}.
Now replacing $\pi$ and $\eta$ by $\pi^*$ and $\eta^*$ in the HJBI equation \eqref{eq:HJBI}, we obtain 
\begin{equation}
	\label{eq:dh}
	\begin{cases}
		0 = h'(t) + r h(t),  & h(T) = 1, \\
		0 = f'(t) + \left(2 \xi^\top \xic - \left\| \xic \right\|^2 \right) f(t), & f(T) = 1,
	\end{cases}
\end{equation}
which lead to the explicit solutions given by \eqref{eq:h_f}. 

In the next step, we show that $\pi^*(\cdot)$ is given by \eqref{eq:pi_MMV}.
However, we cannot directly calculate $\Lambda^*(\cdot):=\Lambda^{\eta^*}(\cdot)$ and substitute it into \eqref{eq:can} to obtain \eqref{eq:pi_MMV}, where  $\Lambda^*(\cdot)$ satisfies the following SDE (using \eqref{eq:dQ}):
\begin{align}
	\label{eq:dLam_op}
	\dd \Lambda^*(t) 
	= - \Lambda^*(t)  \, \xic^\top \, \dd W(t), \quad \Lambda^*(0) =1.
\end{align}
The reason is that \eqref{eq:can} offers a candidate to the optimal strategy $\pi^*(t)$ to the dynamic MMV problems \eqref{eq:Vt} given $(t, X^*(t), \Lambda^*(t)) = (t, x, \lambda)$, but  there is an implicit constraint on the relation between $\Lambda^*(\cdot)$ and  $X^*(\cdot)$. Due to this constraint, $\Lambda^*(t)$ cannot be equal to a free variable $\lambda$ but rather depends on $X^*(t) =x$.\footnote{If such a relation (see \eqref{eq:relation}) were not recognized, one would mistakenly conclude from \eqref{eq:can} that MMV problems are time-consistent. However, upon noticing this relation, we obtain the optimal MMV strategy in \eqref{eq:pi_MMV}, which depends on the initial wealth $x_0$ and immediately shows that MMV problems are time-inconsistent. 
}
To establish the relation between $\Lambda^*(\cdot)$ and  $X^*(\cdot)$, we first apply It\^o's lemma to obtain (using \eqref{eq:dX}, \eqref{eq:can}, \eqref{eq:dh}, and \eqref{eq:dLam_op})
\begin{align*}
	\dd \big( h(t) X^*(t) \big) &= \frac{f(t) \Lambda^*(t)}{\theta} \left(\xi^\top \xic \, \dd t + \xic^\top \, \dd W(t) \right), 
	\\
	\dd \big( f(t) \Lambda^*(t) \big) &= - f(t) \Lambda^*(t) \left( \left(2 \xi^\top \xic - \left\| \xic \right\|^2 \right) \dd t + \xic \, \dd W(t) \right).
\end{align*}
We next introduce a new family of processes $G(\cdot, \cdot) := \{G(t,s)\}_{0 \le t \le s \le T}$  defined by 
\begin{align}
	\label{eq:G}
	G(t, s) := \int_t^s \, g(u) \, \dd u,  \quad \text{with }
	g(t) := f(t) \Lambda^*(t) \left( \left\| \xic \right\|^2 - \xi^\top \xic \right),
\end{align}
where $f(t)$ is given in \eqref{eq:h_f} and $\Lambda^*(t)$ is solved from \eqref{eq:dLam_op}.
By combining the above results, we finally get
\begin{align}
	\label{eq:relation}
	\theta \big( h(s) X^*(s) - h(t) X^*(t) \big) = -  \big( f(s) \Lambda^*(s) - f(t) \Lambda^*(t) \big) +  G(t,s) \qquad \forall \, 0 \le t \le s \le T.
\end{align}
Taking $t = 0$ and $s = t$ in \eqref{eq:relation}, noting $X^*(0) = x_0$ and $\Lambda^*(0) = 1$, and recalling $h$ and $f$ from \eqref{eq:h_f}, we obtain $\pi^*(\cdot)$ in \eqref{eq:pi_MMV}. 

In the remaining part of the proof, we show that all the conditions in Lemma \ref{lem:veri} hold, which is achieved by the following steps:
\begin{itemize}
	\item Show $(\pi^*(\cdot), \eta^*(\cdot)) \in \Ac \times \Bc$. Since we have $\eta^*(t) \equiv - \xic$ for all $t$ from \eqref{eq:can}, $\eta^*(\cdot)$ is a Markov control and $\eta^*(\cdot) \in \Bc$ is  confirmed. 
	To show $\pi^*(\cdot)$ given by \eqref{eq:pi_MMV} is admissible, first notice from \eqref{eq:can} that $\pi^*(t) \in \Pi$ for all $t \in [0,T]$. 
	Next, as $f$, $\Lambda^*(\cdot)$, and $G(0, \cdot)$ are all explicitly given, the existence and uniqueness of $X^*(\cdot)$ to \eqref{eq:dX} under $\pi^*(\cdot)$ is established by the relation \eqref{eq:relation}. Finally, using \eqref{eq:relation} again, along with the boundedness of $f$  and $h$ and the square-integrability of $\Lambda^*(\cdot)$ (thus also $G(0, \cdot)$), we have $\Eb^\Pb [ \sup_{t \in [0,T]} |\Lambda^*(t)|^2 ] \le 4 \Eb^\Pb [  |\Lambda^*(T)|^2 ] < +\infty$ and $\Eb^\Pb [\sup_{t \in [0,T]} |X^*(t)|^2] < \infty$, which in turn implies the square-integrability of $\pi^*(\cdot)$. Therefore, by the definition of admissible strategies on page 4, we conclude $\pi^*(\cdot) \in \Ac$.
	\item By recalling $h$ and $f$ from \eqref{eq:h_f} and $V$ from \eqref{eq:V_ansatz}, we easily see that $V \in \mathrm{C}^{1,2,2}((0,T) \times \Rb \times (0,\infty)) \cap \mathrm{C}([0,T] \times \Rb \times [0,\infty) )$.
	According to the above calculations (in particular, $(\pi^*, \eta^*)$ in \eqref{eq:can} and the ODE system \eqref{eq:dh}), $V$ in \eqref{eq:V_ansatz} satisfies conditions (i)-(iii) in Lemma \ref{lem:veri}.
	Since $h(T)=f(T)=1$, condition (iv) in Lemma \ref{lem:veri} also holds.
	\item $\forall \eta(\cdot) \in \Bc$, the SDE $\dd \Lambda^\eta(s) =  \Lambda^\eta(s) \eta(s)^\top \, \dd W (s)$, with $\Lambda^\eta(t) = \lambda \in \Rb_+$, admits a unique strong solution such that $\Eb^{\Pb}_{t,x,\lambda}[\sup_{s \in [t, T]} |\Lambda^\eta(s)|^2 ] < +\infty$,  which implies 
	that  $\Eb_{t,x,\lambda}^{\Qb^\eta} [\sup_{s \in [t, T]} |\Lambda^\eta(s)|] \le \Eb^\Pb_{t,x,\lambda}[\sup_{s \in [t, T]} |\Lambda^\eta(s)|^2 ] < +\infty$ immediately.
	$\forall \pi(\cdot) \in \Ac$, we have from the definition of admissible strategies that $\Eb^\Pb_{t,x,\lambda}[\sup_{s \in [t, T]} |X(s)|^2 ] < +\infty$. 
	By the Cauchy-Schwarz inequality, it follows that $\Eb_{t,x,\lambda}^{\Qb^\eta} [\sup_{s \in [t, T]} |X(s)|] \le \sqrt{\Eb^{ \Pb}_{t,x,\lambda}[\sup_{s \in [t, T]} |\Lambda^\eta(s)|^2 ]} \sqrt{\Eb^\Pb_{t,x,\lambda}[\sup_{s \in [t, T]} |X(s)|^2 ]} < +\infty$. 
	Now using \eqref{eq:V_ansatz} along with the boundedness of $h$ and $g$ from \eqref{eq:h_f}, we obtain 
	\begin{align*}
		\Eb_{t,x,\lambda}^{\Qb^\eta} \left[\sup_{s \in [t, T]}  \, \big| V(s, X(s), \Lambda^\eta(s)) \big| \right] \le K \left(1 + \Eb_{t,x,\lambda}^{\Qb^\eta} \left[\sup_{s \in [t, T]} |X(s)| + \sup_{s \in [t, T]} |\Lambda^\eta(s)| \right]\right)< + \infty
	\end{align*}
for some positive constant $K$, which confirms condition (v) in Lemma \ref{lem:veri}.
\end{itemize}
Finally, by recalling Lemma \ref{lem:veri} again, all the results in Theorem \ref{thm:MMV} follow immediately.
\end{proof}

\subsection{Solution to Constrained MV Problems}
\label{sub:MV}

Constrained MV problems under a conic convex constraint, as in Problem \ref{prob:MV}, have been studied before in the literature (see, e.g., \cite{li2002dynamic}, \cite{hu2005constrained},  \cite{czichowsky2013cone}, and \cite{li2016continuous}). Here we provide solutions  in Theorem \ref{thm:MV}  with an abbreviated proof, which heavily relies on the embedding technique first proposed in \cite{li2000optimal} and \cite{zhou2000continuous}.

Introduce a family of auxiliary processes $\wt X_\beta(\cdot)$, indexed by $\beta \in \Rb$, which are defined by
\begin{align}
	\label{eq:X_wt}
	\wt X_\beta(t) = \wt X(t) - \beta / h(t) = \wt X(t) - \beta \, \ee^{-r(T-t)}, \qquad t \in [0,T],	
\end{align} 
where $\wt X(\cdot)$ solves the SDE \eqref{eq:dX} under an admissible strategy $\wt \pi(\cdot) \in \Ac$ (in particular, $\wt \pi(t) \in \Pi$ for all $t \in [0,T]$) and $h$ is defined in \eqref{eq:h_f}. 
Using \eqref{eq:dX} and \eqref{eq:X_wt}, we obtain the dynamics of $\wt X_\beta(\cdot)$ by
\begin{align}
	\label{eq:dX_wt}
	\dd \wt X_\beta(t) = \left( r \wt X_\beta(t) + \wt \pi(t)^\top B \right) \, \dd t + \wt \pi(t)^\top \sigma \, \dd W(t), \qquad \wt X_\beta(0) = x_0 - \beta \, \ee^{-rT}.
\end{align}
Consider the following quadratic loss minimization problem:
\begin{align}
	\label{eq:QLM}
	V_{\mathrm{QLM}}(t, x) := \inf_{\wt \pi(\cdot) \in \Ac} \, \Eb_{t,x} \left[ \big(\wt X_\beta (T) \big)^2 \right], \quad t \in [0,T], \, x \in \Rb, 
\end{align}
where $\wt X_\beta (T)$ is the terminal value of $\wt X_\beta(\cdot)$ in \eqref{eq:dX_wt}, starting from $\wt X_\beta(t) = x$. 
The following lemma links Problem \eqref{eq:QLM} with Problem \ref{prob:MV} (constrained MV problems). 
We omit its proof here, since a similar  proof can be found in \cite{zhou2000continuous}.

\begin{lemma}
	\label{lem:MV}
	Let $\wt \pi^*(\cdot)$ 
	denote the optimal strategy to Problem \ref{prob:MV} and $\wt X^*(\cdot)$ the corresponding optimal wealth process; for any $\beta \in \Rb$, let $\wt \pi^*_\beta(\cdot)$ denote the optimal strategy to Problem \eqref{eq:QLM}  and $\wt X^*_\beta(\cdot)$ the corresponding optimal wealth process, both under the initial condition $(t,x) = (0, x_0)$. 
	Then $\wt \pi^*(\cdot) =\wt \pi^*_{\beta^*} (\cdot) $, where $\beta^*$ is the unique solution to the following equation of $\beta$:
	\begin{align}
		\label{eq:beta}
		\beta = \frac{1}{\theta} + \Eb \left[ \wt X_\beta^*(T)  + \beta \right] = \frac{1}{\theta} + \Eb \big[ \wt X^*(T)\big].
	\end{align}  
\end{lemma}

\begin{theorem}
	\label{thm:MV}
	The optimal investment strategy $\wt \pi^*(\cdot)$ to Problem \ref{prob:MV} (constrained MV problems) is given by 
	\begin{align}
		\label{eq:pi_MV}
		\wt \pi^*(t) = - \left(\wt X^*(t) - x_0 \, \ee^{rt} - \frac{1}{\theta} \, \ee^{-r(T-t) + \xi^\top \xic T}\right) \Sigma^{-1} \sigma \xic, \qquad t \in [0,T],
	\end{align}
	where $\wt X^*(\cdot)$ is the corresponding optimal wealth process \eqref{eq:dX} under the above $\wt \pi^*(\cdot)$ and $\wt X^*(0) = x_0$.
	
\end{theorem}

\begin{proof}
	We first solve Problem \eqref{eq:QLM} under an arbitrary but fixed initial condition $(t,x)$. 
	To that end, notice that the value function $\VQ$ satisfies the following Hamilton-Jacobi-Bellman equation:
	\begin{align}
		\label{eq:VQ}
		0 = \inf_{\wt \pi \in \Pi} \, \left\{\frac{\partial \VQ}{\partial t} + \frac{\partial \VQ}{\partial x} \big[rx + \wt \pi^\top B \big] + \frac{1}{2} \frac{\partial^2 \VQ}{\partial x^2} \, \wt \pi^\top \Sigma \wt \pi \right\}
	\end{align}
	and the boundary condition $\VQ(T, x) = x^2$.  We guess an {\it ansatz} of $\VQ$ in the form of 
	\begin{align}
		\label{eq:VQ_sol}
		\VQ(t,x) = x^2 \, \wt f(t), \qquad \text{with } \wt f(t)>0 \text{ and } \wt f(T) = 1.
	\end{align}
	Since $\wt f(t) >0$, the first-order condition is also sufficient and yields the optimizer as  
	\begin{align}
		\label{eq:pi_QLM}
		\wt \pi^*_\beta = - x \, \Sigma^{-1} \sigma \xic ,
	\end{align}
	where $\xic$ is defined in \eqref{eq:xi}. Plugging the above $\wt \pi^*_\beta$ into \eqref{eq:VQ}, we obtain
	\begin{align*}
		\wt f'(t) + \left(2r - 2 \xi^\top \xic + \left\| \xic \right\|^2 \right) \wt f(t) = 0 ,
	\end{align*}
	which, along with $\wt f(T) =1$, gives 
	\begin{align}
		\label{eq:f_wt}
		\wt f(t) = \ee^{ \left( 2r - 2 \xi^\top \xic + \left\| \xic \right\|^2 \right) \, (T - t)} . 
	\end{align}

Next, we follow \cite{fleming2006controlled}[Theorem IV.3.1., p.157] to verify that $\VQ$ in \eqref{eq:VQ_sol} is the value function and $\wt \pi^*_\beta(\cdot)$ in \eqref{eq:pi_QLM} is the optimal strategy of the QLM problem \eqref{eq:QLM}. 
By \eqref{eq:f_wt}, $\wt f$ is differentiable and bounded, and thus $\VQ(t,x) = x^2 \, \wt f(t) \in \mathrm{C}^{1,2}((0,T) \times \Rb) \cap \mathrm{C}([0,T] \times \Rb)$ and satisfies a polynomial growth condition with $p \ge 2$. The above computations confirm that  $\VQ$ is a classical solution to the HJB equation \eqref{eq:VQ}, satisfying the boundary condition.
Furthermore, $\wt \pi^*_\beta(\cdot)$ in \eqref{eq:pi_QLM} is a Markov control, leads to a unique solution $\wt X^*_\beta(\cdot)$ of \eqref{eq:dX_wt} (see below), and satisfies the square-integrability condition since $\Eb^\Pb[\sup_{t \in [0,T]} \, |\wt X^*_\beta(t)|^2] < \infty$ by the standard theory on SDEs.

	With $\wt \pi(\cdot)$ replaced by $\wt \pi_\beta^*$ from \eqref{eq:pi_QLM}, we easily solve the SDE \eqref{eq:dX_wt} and obtain $\wt X^*_\beta(\cdot)$ by 
	\begin{align*}
		\wt X_\beta^*(t) = \left(x_0 - \beta \, \ee^{-r T} \right) \, \ee^{ \big(r - \xi^\top \xic - \frac{1}{2} \left\| \xic \right\|^2  \big) t \, - \,  \xic^\top W(t)}, \quad t \in [0,T].
	\end{align*}
	Combining \eqref{eq:beta} with the above, we find the unique $\beta^*$ by 
	\begin{align*}
		\beta^* = x_0 \, \ee^{rT} + \frac{1}{\theta} \, \ee^{\xi^\top \xic T} ,
	\end{align*}
	which, in combination with \eqref{eq:X_wt} and \eqref{eq:pi_QLM} and Lemma \ref{lem:MV}, proves \eqref{eq:pi_MV}.  
\end{proof}

\subsection{Comparison Results}
\label{sub:com}

With explicit optimal strategies to both MMV problems and MV problems found in previous sections, we carry out a comparison study to investigate whether they remain the same in our formulation. 
In both the unconstrained case and the constrained case, Corollaries \ref{cor:un} and \ref{cor:con} show that the two optimal strategies $\pi^*(\cdot)$ in \eqref{eq:pi_MMV} and $\wt \pi^*(\cdot)$ in \eqref{eq:pi_MV}  coincide. 

\subsubsection{The Unconstrained Case}

Since $\Rb^n$ itself is convex and conic, we consider in this section the unconstrained case of $\Pi = \Rb^n$, or equivalently, $\Pis = \Rb^d$. 
Recall the definitions of $\xi$ and $\xic$ in \eqref{eq:xi}, we now have $\xic = \xi$, 
which immediately implies that 
\begin{align*}
	f(t) = \ee^{\xi^\top \xi (T - t)} = \ee^{B^\top \Sigma^{-1} B (T - t)} \quad \text{ and } \quad G(t, s) \equiv 0, \quad  0 \le t \le s \le T,
\end{align*}
where $f$ and $G$ are defined respectively by  \eqref{eq:h_f} and \eqref{eq:G}. 
By Theorems \ref{thm:MMV} and \ref{thm:MV}  (in particular, \eqref{eq:pi_MMV} and \eqref{eq:pi_MV}), 
we obtain the following corollary:

\begin{corollary}
	\label{cor:un}
Suppose $\Pi = \Rb^n$. MMV problems and MV problems, with the same risk aversion parameter $\theta$ and the same initial 
wealth $x_0$, have the same optimal investment strategy given by 
\begin{align}
	\label{eq:pi_un}
	\wt \pi^*(t) = \pi^*(t) =  - \Bigg(X^*(t) - \underbrace{ \left( x_0 \, \ee^{rt} + \frac{1}{\theta} \, \ee^{-r(T-t) + B^\top \Sigma^{-1} B T} \right)}_{:=\chi(t), \, \text{a dynamic threshold}} \Bigg) \Sigma^{-1} B, \quad t \in [0,T].
\end{align}
\end{corollary}

Although the financial market introduced in Section \ref{sec:model} may be incomplete (whenever $d > n$), the price processes of the risky asset are continuous, which is a special case of \cite{strub2020note} and \cite{vcerny2020semimartingale}. 
As such, it is not surprising that the result in Corollary \ref{cor:un} coincides with theirs. 
In fact, with $G(t, s) \equiv 0$, the relation \eqref{eq:relation} reduces to 
\begin{align*}
	\theta \big( h(s) X^*(s) - h(t) X^*(t) \big) = -  \big( f(s) \Lambda^*(s) - f(t) \Lambda^*(t) \big), \qquad \forall \, 0 
	\le t \le s \le T.
\end{align*}
Then, we can carry out some simple computations to get 
\begin{align*}
	X^*(T) - \Eb[X^*(T)] = \frac{1}{\theta} (1 - \Lambda^*(T))\le \frac{1}{\theta}.
\end{align*}
Such an inequality implies that the optimal terminal wealth of both the unconstrained MMV and MV problems falls in the 
monotone region $\mathcal{G}_\theta$,
where MMV preferences coincide with MV preferences (see \cite{maccheroni2009portfolio} and \cite{vcerny2012computation}), justifying the result of Corollary \ref{cor:un}.

\subsubsection{The Constrained Case}
\label{subsub:con}

We now focus on the constrained case, where the conic convex set $\Pi$ is a true subset of $\Rb^n$. By a careful examination 
of $\pi^*(\cdot)$ in \eqref{eq:pi_MMV} and $\wt \pi^*(\cdot)$ in \eqref{eq:pi_MV}, both share the same form as follows:
\begin{align}
	\label{eq:form}
	\text{optimal strategy at } t = - \Bigg(\text{wealth at } t - 
	\underbrace{\left( x_0 \ee^{rt} + \frac{1}{\theta} \ee^{-r(T-t)} \, \text{``factor''} \right)}_{:= \chi_c(t), \, \text{a dynamic threshold}} \Bigg) \times \Sigma^{-1} \sigma \xic,
\end{align}
where the ``factor'' is the only seemingly different component and is given by 
\begin{equation}
	\label{eq:factor}
\text{``factor''} =	
    \begin{cases}
	\Psi(t) := f(0) + G(0,t),   & \text{ (MMV problems)} \\
	\wt \Psi := \ee^{\xi^\top \xic T},   & \text{ (MV problems)}
	\end{cases}.
\end{equation}
Regarding $\Psi(t)$ in \eqref{eq:factor}, $f(0) = \ee^{( 2 \xi^\top \xic - \left\| \xic \right\|^2 ) T}$ is a constant by its definition \eqref{eq:h_f}, but $G(0,t)$ is an $\Fc_t$-measurable random variable, according to \eqref{eq:G}.
This easily leads to an impression that $\Psi(t) \neq \wt \Psi$ for some $t$ and thus $\pi^*(\cdot) \neq \wt \pi^*(\cdot)$ in general.
However, as will become clear later, this impression is wrong! 
Let us resort to a simple example for insight.

\begin{example}
	\label{exm:one}
Consider a simple one-dimensional example with $n = d = 1$ and $\Pi = [0, \infty)$ (no short-selling constraint). 
(1) Assume $\mu < r$. 
By \eqref{eq:xi}, $\xi = (\mu - r)/\sigma < 0 \neq \xic = 0$. 
Hence, according to \eqref{eq:pi_MMV} and \eqref{eq:pi_MV}, $\pi^*(t) = \wt \pi^*(t) = 0$ for all $t \in [0,T]$.
(2) Assume $\mu \ge r$. 
We have $\xi = \xic \ge 0$, which implies $g(t) \equiv 0$  and thus $G(0, t)=0$ for all $t$ by \eqref{eq:G}. Using \eqref{eq:h_f}, $\Psi(t) = f(0) + G(0,t) = f(0) = \ee^{\xi^2 T} = \wt \Psi$. Therefore, we have $\pi^*(t) = \wt \pi^*(t)$ for all $t \in [0,T]$. 
\end{example}

Observe from the above example that $\xi \xic - \xic^2 = 0$ holds true in both cases. 
The lemma below generalizes this observation to any conic convex set.

\begin{lemma}
	\label{lem:xi}
Let $\Pi$ be a conic convex subset of $\Rb^n$, and recall $\Pis = \{\sigma^\top \pi : \pi \in \Pi \}$,  $\xi$ and $\xic$ are defined in \eqref{eq:xi}. We always have 
\begin{align}
	\label{eq:rela_xi}
	\xi^\top \xic - \left\| \xic \right\|^2 = 0.
\end{align}
\end{lemma}

\begin{proof}
	Recall the second projection theorem  (see Theorem 9.8 in \cite{beck2014introduction}) states that $\xic$ is the orthogonal projection of $\xi$ onto 
	$\Pi_\sigma$
	if and only if the following relation holds
$(\xi - \xi_c)^\top (z - \xi_c) \leq 0$,  $\forall z \in \Pi_\sigma$. 
On the one hand, by setting $z = 0_d \in \Pi_\sigma$, we obtain 
$	(\xi - \xi_c)^\top \xi_c \geq 0 $.
On the other hand, since the constraint set $\Pi_\sigma$ is conic, if $\xi_c \in \Pi_\sigma$, then $2 \xi_c \in \Pi_\sigma$.
Thereby, taking $z = 2 \xi_c$ gives 
$	(\xi - \xi_c)^\top \xi_c \leq 0 $.
Therefore, the desired result \eqref{eq:rela_xi} follows immediately.
\end{proof}

\begin{remark}\label{rmk:general}
	Lemma \ref{lem:xi} plays a key role in guaranteeing that the constrained MMV and MV problems have the same optimal solution under 
	the conic convex constraint. The relation \eqref{eq:rela_xi} may fail for a general convex set. For instance, 
	if we focus on $\mathbb R$ and take $\Pi_\sigma = [0, 1]$ and $\xi = 2$, then $\xi_c = 1$ and $\xi \xi_c - \xi_c^2
	= 1 \neq 0$.
\end{remark}

Now using \eqref{eq:rela_xi}, along with \eqref{eq:h_f} and \eqref{eq:G}, we easily see that 
\begin{align*}
	f(0) = \ee^{\xi^\top \xic T} \quad \text{ and } \quad 
	g(t) \equiv 0, \; G(0,t) = \int_0^t \, g(s) \, \dd s = 0, 
\end{align*}
which shows $\Psi(t) =\wt \Psi$ for all $ t \in [0,T]$ and the following corollary: 

\begin{corollary}
	\label{cor:con}
	Constrained MMV problems and constrained MV problems, with the same risk aversion parameter $\theta$ and the same initial wealth $x_0$, have the same optimal investment strategy given by 
	\begin{align}
		\label{eq:pi_con}
		\wt \pi^*(t) = \pi^*(t) =  - \left(X^*(t) - x_0 \, \ee^{rt} - \frac{1}{\theta} \, \ee^{-r(T-t) + \xi^\top \xic T}\right) \Sigma^{-1} B, \quad t \in [0,T].
	\end{align}
\end{corollary}

\begin{remark}
Note that $B^\top \Sigma^{-1} B = \xi^\top \xi$ in \eqref{eq:pi_un} (the unconstrained case), which is in general not equal to $\xi^\top \xic$ in \eqref{eq:pi_con} (the constrained case). See Case (1) of Example \ref{exm:one}: $\xi^\top \xi > 0$ but $\xi^\top \xic = 0$.
\end{remark}

\subsubsection{Further Discussions}
\label{subsub:dis}

We close this section by carrying out further discussions on the equivalence result between MMV problems and MV problems obtained in Corollaries \ref{cor:un} and \ref{cor:con}.

Let us first focus on the unconstrained case. \cite{maccheroni2009portfolio} comment that MMV agents may still be regarded using the MV functional $\wt J_\theta$ \eqref{eq:MV} when evaluating prospects $X$ outside the monotone region $\mathcal{G}_\theta$, but in such a case, a truncated version $X \wedge t$, \emph{not} $X$ itself, is evaluated by $\wt J_\theta$, where $t$ is the largest constant such that  $X \wedge t \in \mathcal{G}_\theta$. 
Theorem 4.1 therein further proves that the optimal terminal wealth to Problem \ref{prob:MMV} is obtained from $\Eb^\Pb[( X^*(T) - \kappa^*)^-] = 1/\theta$ for a unique threshold $\kappa^* \in \Rb$. 
By Theorem \ref{thm:MMV}, straightforward computations yield
$\Eb^\Pb[ X^*(T)] + 1/\theta= \chi(T)$,
where $\chi(\cdot)$ is  defined in \eqref{eq:pi_un}.
We next argue that $ X^*(t) \le \chi(t)$ holds for all $t \in [0,T]$. 
To see this, notice that $ X^*(0) = x_0 < \chi(0)$ and 
if $ X^*(\tau) = \chi(\tau)$ at some time $\tau \le T$, the investment in the risky asset(s) $\pi^*(\cdot)$ becomes 0 at $\tau$  by \eqref{eq:pi_un} and will stay at 0 afterwards, since $\dd \chi(t) = r \chi(t) \, \dd t$ and $X^*(\cdot)$ grows deterministically  at the same rate $r$ over $[\tau, \tau + \dd t]$. (This implies $\kappa^* = \chi(T)$ and explains  why we call $\chi(\cdot)$ the the dynamic threshold process.)
We remark that both price continuity and dynamic trading are essential to $X^*(\cdot) \le \chi(\cdot)$; price continuity ensures that the optimal portfolio process $X^*(\cdot)$ does not jump over the dynamic threshold $\chi(\cdot)$, while dynamic trading allows agents to stop at exactly the time when the two processes meet.
The above explanations offer a complete ``picture'' of the equivalence between MMV problems and MV problems for the unconstrained case.

We next understand the above result from an alternative angle.
\cite{vcerny2020semimartingale} provides a nice commutative diagram (see (7) therein) showing that there are two equivalent routes of arriving at MMV preferences:
\begin{enumerate}
\item[(1)] a quadratic utility $U_\theta(x) = x - \frac{\theta}{2} x^2$ $\stackrel{\square C}{\longrightarrow}$ MV preferences \eqref{eq:MV} $\stackrel{\square D}{\longrightarrow}$ MMV preferences \eqref{eq:MMV}, 
\item[(2)] $U_\theta(x)$ $\stackrel{\square D}{\longrightarrow}$ a \emph{truncated} quadratic utility $\wt U_\theta(x) = U_\theta(x \wedge 1/\theta) $  $\stackrel{\square C}{\longrightarrow}$ MMV preferences \eqref{eq:MMV},
\end{enumerate}
where $\square$ denotes the supremal convolution operator, and $C$ (resp., $D$) is a concave cash (resp., positive cone) indicator function. 
($\square C$ may be understood as ``cash invariance'' while $\square D$ yields monotonization.)
The first route is taken by \cite{maccheroni2009portfolio} in defining MMV preferences. 
Utilizing the second route, \cite{vcerny2020semimartingale} shows in Theorem 5.4 that the optimal wealth process of MMV problems coincides with that of MV problems if and only if the latter does not go beyond the ``bliss point'', which holds true for continuous price processes (see \cite{delbaen1996variance}).

Moving to the constrained case, we rely on the intuition from the unconstrained case and investigate how constraints affect optimal strategies. As noted after  \eqref{eq:Proj}, the \emph{conic} assumption  implies that the projection $\Proj_{\Pis}$ is homogeneous, satisfying  $\Proj_{\Pis} [k v] = k \, \Proj_{\Pis} [v]$, which is key to deriving the optimal strategies $\pi^*(\cdot)$ in \eqref{eq:pi_MMV} for Problem \ref{prob:MMV} and $\wt \pi^*(\cdot)$ in \eqref{eq:pi_MV} for Problem \ref{prob:MV}. 
We then observe that both $\pi^*(\cdot)$ and $\wt \pi^*(\cdot)$ share the same form \eqref{eq:form}, and constraints change the threshold process from $\chi(\cdot)$ in \eqref{eq:pi_un} to $\chi_c(\cdot)$ in \eqref{eq:form}. 
At this moment, we should introduce two threshold processes, one for $X^*(\cdot)$ from Problem \ref{prob:MMV} and the other for $\wt X^*(\cdot)$ from Problem \ref{prob:MV}; however, the orthogonality condition \eqref{eq:rela_xi} implies the two  $\chi_c(\cdot)$ defined in \eqref{eq:form} coincide in the two problems.
Moreover, $\chi_c(\cdot)$ follows the same dynamics as $\chi(\cdot)$ given by  $\dd \chi_c(t) = r \chi_c(t) \, \dd t$ and satisfies $x_0 < \chi_c(0)$. 
Therefore, by following similar arguments in the unconstrained case, we conclude that $\chi_c(T)$ is the ``bliss point'' and both $X^*(\cdot)$ and $\wt X^*(\cdot)$ are dominated by $\chi_c(\cdot)$ in the constrained case. 
As a result, the equivalence result in Corollary \ref{cor:con} follows.

\section{Conclusions}
\label{sec:con}

We solve constrained MMV problems and constrained MV problems in a standard GBM market model, where investment strategies are subject to a conic convex constraint. 
A careful examination shows that the optimal strategies to these two problems coincide, without and with the constraint. This finding then partially extends the existing equivalence result between MMV and MV preferences to a constrained setup, in a specific GBM model. 
It would be interesting in future research to investigate whether this equivalence still holds in general market models, or under convex, but not necessarily conic, constraints. 
The latter is challenging but could be rewarding, as the key relation \eqref{eq:rela_xi} in Lemma \ref{lem:xi} leading to  Corollary  \ref{cor:con}  holds under a conic convex constraint but can easily fail under a general non-conic one. 

\vspace{2ex}
\noindent
\textbf{Acknowledgment.} We would like to thank  Ale{\v{s}} {\v{C}}ern{\`y}  for suggesting us consider constrained MMV problems 
in a simple setting (the GBM model) and for discussions on time-inconsistency of MMV and MV problems, and on the economic meaning of the mathematical results in Section \ref{subsub:dis}. 
We are also grateful to the comments from two anonymous reviewers and an associate editor.
YS acknowledges partial support from an ARC Discovery Early Career Research Award (Grant No. DE200101266) and BZ acknowledges partial support by a start-up grant from the University of Connecticut.

\singlespacing

\bibliographystyle{apalike}
\bibliography{reference}

\begin{thebibliography}{}

\bibitem[Basak and Chabakauri, 2010]{basak2010dynamic}
Basak, S. and Chabakauri, G. (2010).
\newblock Dynamic mean-variance asset allocation.
\newblock {\em Review of Financial Studies}, 23(8):2970--3016.

\bibitem[B{\"a}uerle and Grether, 2015]{bauerle2015complete}
B{\"a}uerle, N. and Grether, S. (2015).
\newblock Complete markets do not allow free cash flow streams.
\newblock {\em Mathematical Methods of Operations Research}, 81(2):137--146.

\bibitem[Beck, 2014]{beck2014introduction}
Beck, A. (2014).
\newblock {\em Introduction to Nonlinear Optimization: Theory, Algorithms, and
  Applications with MATLAB}.
\newblock Society for Industrial and Applied Mathematics.

\bibitem[Bj{\"o}rk and Murgoci, 2010]{bjork2010general}
Bj{\"o}rk, T. and Murgoci, A. (2010).
\newblock A general theory of {Markovian} time inconsistent stochastic control
  problems.
\newblock {\em Available at SSRN 1694759}.

\bibitem[{\v{C}}ern{\`y}, 2020]{vcerny2020semimartingale}
{\v{C}}ern{\`y}, A. (2020).
\newblock Semimartingale theory of monotone mean--variance portfolio
  allocation.
\newblock {\em Mathematical Finance}, 30(3):1168--1178.

\bibitem[{\v{C}}ern{\`y} et~al., 2012]{vcerny2012computation}
{\v{C}}ern{\`y}, A., Maccheroni, F., Marinacci, M., and Rustichini, A. (2012).
\newblock On the computation of optimal monotone mean--variance portfolios via
  truncated quadratic utility.
\newblock {\em Journal of Mathematical Economics}, 48(6):386--395.

\bibitem[Cui et~al., 2017]{cui2017mean}
Cui, X., Li, D., and Li, X. (2017).
\newblock Mean-variance policy for discrete-time cone-constrained markets:
  {Time} consistency in efficiency and the minimum-variance signed
  supermartingale measure.
\newblock {\em Mathematical Finance}, 27(2):471--504.

\bibitem[Cui et~al., 2012]{cui2012better}
Cui, X., Li, D., Wang, S., and Zhu, S. (2012).
\newblock Better than dynamic mean-variance: {Time} inconsistency and free cash
  flow stream.
\newblock {\em Mathematical Finance}, 22(2):346--378.

\bibitem[Czichowsky and Schweizer, 2013]{czichowsky2013cone}
Czichowsky, C. and Schweizer, M. (2013).
\newblock Cone-constrained continuous-time markowitz problems.
\newblock {\em Annals of Applied Probability}, 23(2):764--810.

\bibitem[Delbaen and Schachermayer, 1996]{delbaen1996variance}
Delbaen, F. and Schachermayer, W. (1996).
\newblock The variance-optimal martingale measure for continuous processes.
\newblock {\em Bernoulli}, 2(1):81--105.

\bibitem[Fan et~al., 2022]{fan2022monotone}
Fan, K., Shen, Y., Wei, J., and Zou, B. (2022).
\newblock Monotone mean-variance portfolio selection with convex constraints
  under {non-Markovian} regime-switching models.
\newblock {\em Working Paper}.

\bibitem[Fleming and Soner, 2006]{fleming2006controlled}
Fleming, W.~H. and Soner, H.~M. (2006).
\newblock {\em Controlled Markov Processes and Viscosity Solutions}, volume~25.
\newblock Springer Science \& Business Media.

\bibitem[Hansen and Sargent, 2001]{hansen2001robust}
Hansen, L.~P. and Sargent, T.~J. (2001).
\newblock Robust control and model uncertainty.
\newblock {\em American Economic Review}, 91(2):60--66.

\bibitem[Hu and Zhou, 2005]{hu2005constrained}
Hu, Y. and Zhou, X.~Y. (2005).
\newblock Constrained stochastic {LQ} control with random coefficients, and
  application to portfolio selection.
\newblock {\em SIAM Journal on Control and Optimization}, 44(2):444--466.

\bibitem[Li and Ng, 2000]{li2000optimal}
Li, D. and Ng, W.-L. (2000).
\newblock Optimal dynamic portfolio selection: Multiperiod mean-variance
  formulation.
\newblock {\em Mathematical Finance}, 10(3):387--406.

\bibitem[Li and Xu, 2016]{li2016continuous}
Li, X. and Xu, Z.~Q. (2016).
\newblock Continuous-time {Markowitz’s} model with constraints on wealth and
  portfolio.
\newblock {\em Operations Research Letters}, 44(6):729--736.

\bibitem[Li et~al., 2002]{li2002dynamic}
Li, X., Zhou, X.~Y., and Lim, A.~E. (2002).
\newblock Dynamic mean-variance portfolio selection with no-shorting
  constraints.
\newblock {\em SIAM Journal on Control and Optimization}, 40(5):1540--1555.

\bibitem[Maccheroni et~al., 2009]{maccheroni2009portfolio}
Maccheroni, F., Marinacci, M., Rustichini, A., and Taboga, M. (2009).
\newblock Portfolio selection with monotone mean-variance preferences.
\newblock {\em Mathematical Finance}, 19(3):487--521.

\bibitem[Markowitz, 1952]{markowitz1952portfolio}
Markowitz, H. (1952).
\newblock Portfolio selection.
\newblock {\em Journal of Finance}, 7(1):77--91.

\bibitem[Mataramvura and {\O}ksendal, 2008]{mataramvura2008risk}
Mataramvura, S. and {\O}ksendal, B. (2008).
\newblock Risk minimizing portfolios and {HJBI} equations for stochastic
  differential games.
\newblock {\em Stochastics}, 80(4):317--337.

\bibitem[Shen and Zou, 2021]{shen2021mean}
Shen, Y. and Zou, B. (2021).
\newblock Mean--variance investment and risk control strategies--{A}
  time-consistent approach via a forward auxiliary process.
\newblock {\em Insurance: Mathematics and Economics}, 97:68--80.

\bibitem[Strub and Li, 2020]{strub2020note}
Strub, M.~S. and Li, D. (2020).
\newblock A note on monotone mean--variance preferences for continuous
  processes.
\newblock {\em Operations Research Letters}, 48(4):397--400.

\bibitem[Trybu{\l}a and Zawisza, 2019]{trybula2019continuous}
Trybu{\l}a, J. and Zawisza, D. (2019).
\newblock Continuous-time portfolio choice under monotone mean-variance
  preferences—stochastic factor case.
\newblock {\em Mathematics of Operations Research}, 44(3):966--987.

\bibitem[Zhou and Li, 2000]{zhou2000continuous}
Zhou, X.~Y. and Li, D. (2000).
\newblock Continuous-time mean-variance portfolio selection: A stochastic {LQ}
  framework.
\newblock {\em Applied Mathematics and Optimization}, 42(1):19--33.

\end{thebibliography}

\end{document}